\documentclass[onecollarge]{svjour2}       % onecolumn "king-size"
\smartqed  % flush right qed marks, e.g. at end of proof
\usepackage{graphicx}
\newcommand\gothfamily{\usefont{U}{ygoth}{m}{n}}
\DeclareTextFontCommand{\textgoth}{\gothfamily}
\begin{document}

\title{Covariant differentiation of spinors for a general affine connection%\thanks{Grants or other notes
}


\author{Nikodem J. Pop\l awski         %\and
        %Second Author %etc.
}


\institute{N. J. Pop\l awski \at
              Department of Physics, Indiana University, Swain Hall West 117, 727 East Third Street, Bloomington, Indiana 47405, USA \\
              \email{nipoplaw@indiana.edu}           %  \\
           %\and
           %S. Author \at
              %second address
}


\maketitle

\begin{abstract}
We show that the covariant derivative of a spinor for a general affine connection, not restricted to be metric compatible, is given by the Fock--Ivanenko coefficients with the antisymmetric part of the Lorentz connection.
The projective invariance of the spinor connection allows to introduce gauge fields interacting with spinors.
We also derive the relation between the curvature spinor and the curvature tensor for a general connection.
\end{abstract}

\section{Introduction}

Einstein's relativistic theory of gravitation (general relativity) gives a geometrical interpretation of the gravitational field.
The geometry of general relativity is that of a four-dimensional Riemannian manifold, i.e. equipped with a symmetric metric-tensor field and an affine connection that is torsionless and metric compatible.
In the Einstein--Palatini formulation of gravitation~\cite{Pal}, which is dynamically equivalent to the standard Einstein--Hilbert formulation~\cite{FK}, field equations are derived by varying a total action for the gravitational field and matter with respect to the metric tensor and the connection, regarded as independent variables.
The corresponding Lagrangian density for the gravitational field is linear in the symmetric part of the Ricci tensor of the connection.

The postulates of symmetry and metric compatibility of the affine connection determine the connection in terms of the metric.
Relaxing of the postulate of symmetry leads to relativistic theories of gravitation with torsion~\cite{Car,Hehl}.
Relaxing of the postulate of metric compatibility leads to {\em metric--affine} formulations of gravity.
Theories of gravity that incorporate spinor fields use a tetrad as a dynamical variable instead of the metric.
Accordingly, the variation with respect to the affine connection can be replaced by the variation with respect to the Lorentz (spin, anholonomic) connection.
Dynamical variables in relativistic theories of gravitation with relaxed constraints are: metric and torsion (Einstein--Cartan theory)~\cite{Car,Hehl,SG}, metric and asymmetric connection~\cite{HK,HLS}, metric, torsion and nonmetricity~\cite{Smal}, tetrad and torsion~\cite{HD}, and tetrad and spin connection (Einstein--Cartan--Kibble--Sciama theory)~\cite{KS}.

Fock and Ivanenko generalized the relativistic Dirac equation for the electron by introducing the covariant derivative of a spinor in a Riemannian spacetime~\cite{FoIv}.
In order to incorporate spinor fields into a metric--affine theory of gravitation we must construct a covariant differentiation of spinors for a general affine connection.
An example of a physical theory with such a connection is the generalized Einstein--Maxwell theory with the electromagnetic field tensor represented by the second Ricci tensor (the homothetic curvature tensor)~\cite{spinor1}.
In this paper we present a derivation of the covariant derivative of a spinor for a general connection.
We show how the projective invariance of the spinor connection allows to introduce gauge fields interacting with spinors in curved spacetime.
We also derive the formula for the curvature spinor in the presence of a general connection.

\section{Tetrads}

In order to construct a generally covariant Dirac equation, we must regard the components of a spinor as invariant under coordinate transformations~\cite{Schr}.
In addition to the coordinate systems, at each spacetime point we set up four orthogonal vectors (a {\em tetrad}) and the spinor gives a representation for the Lorentz transformations that rotate the tetrad~\cite{Lord,BJ}.
Any vector $V$ can be specified by its components $V^\mu$ with respect to the coordinate system or by the coordinate-invariant projections $V^a$ of the vector onto the tetrad field $e^a_\mu$:
\begin{equation}
V^a=e^a_\mu V^\mu,\,\,\,\,V^\mu=e^\mu_a V^a,
\end{equation}
where the tetrad field $e^\mu_a$ is inverse of $e^a_\mu$:
\begin{equation}
e^\mu_a e^a_\nu=\delta^\mu_\nu,\,\,\,\,e^\mu_a e^b_\mu=\delta^b_a.
\end{equation}
The {\em metric tensor} $g_{\mu\nu}$ of general relativity is related to the coordinate-invariant metric tensor of special relativity $\eta_{ab}=\mbox{diag}(1,-1,-1,-1)$ through the tetrad:
\begin{equation}
g_{\mu\nu}=e^a_\mu e^b_\nu \eta_{ab}.
\label{metric}
\end{equation}
Accordingly, the determinant $\textgoth{g}$ of the metric tensor $g_{\mu\nu}$ is related to the determinant $\textgoth{e}$ of the tetrad $e^a_\mu$ by
\begin{equation}
\sqrt{-\textgoth{g}}=\textgoth{e}.
\end{equation}
We can use $g_{\mu\nu}$ and its inverse $g^{\mu\nu}$ to lower and raise coordinate-based indices, and $\eta_{ab}$ and its inverse $\eta^{ab}$ to lower and raise coordinate-invariant (Lorentz) indices.

Eq.~(\ref{metric}) imposes 10 constraints on the 16 components of the tetrad, leaving 6 components arbitrary.
If we change from one tetrad $e^\mu_a$ to another, $\tilde{e}^\mu_b$, then the vectors of the new tetrad are linear combinations of the vectors of the old tetrad:
\begin{equation}
\tilde{e}^\mu_a=\Lambda^b_{\phantom{b}a}e^\mu_b.
\label{rotation}
\end{equation}
Eq.~(\ref{metric}) applied to the tetrad field $\tilde{e}^\mu_b$ imposes on the matrix $\Lambda$ the orthogonality condition:
\begin{equation}
\Lambda^c_{\phantom{c}a}\Lambda^d_{\phantom{d}b}\eta_{cd}=\eta_{ab},
\label{ortho}
\end{equation}
so $\Lambda$ is a Lorentz matrix.
Consequently, the Lorentz group can be regarded as the group of {\em tetrad rotations} in general relativity~\cite{Hehl,Lord}.

\section{Spinors}

Let $\gamma^a$ be the coordinate-invariant {\em Dirac matrices}:
\begin{equation}
\gamma^a\gamma^b+\gamma^b\gamma^a=2\eta^{ab}.
\label{anticom1}
\end{equation}
Accordingly, the spacetime-dependent Dirac matrices, $\gamma^\mu=e^\mu_a \gamma^a$, satisfy
\begin{equation}
\gamma^\mu\gamma^\nu+\gamma^\nu\gamma^\mu=2g^{\mu\nu}.
\label{anticom2}
\end{equation}
Let $L$ be the spinor representation of a tetrad rotation~(\ref{rotation}):
\begin{equation}
\tilde{\gamma}^a=\Lambda^a_{\phantom{a}b}L\gamma^b L^{-1}.
\label{spinor}
\end{equation}
Since the Dirac matrices $\gamma^a$ are constant in some chosen representation, the condition $\tilde{\gamma}^a=\gamma^a$ gives the matrix $L$ as a function of $\Lambda^a_{\phantom{a}b}$~\cite{Lord,BJ}.
For infinitesimal Lorentz transformations:
\begin{equation}
\Lambda^a_{\phantom{a}b}=\delta^a_b+\epsilon^a_{\phantom{a}b},
\end{equation}
where the antisymmetry of the 6 infinitesimal Lorentz coefficients, $\epsilon_{ab}=-\epsilon_{ba}$, follows from Eq.~(\ref{ortho}), the solution for $L$ is:
\begin{equation}
L=1+\frac{1}{2}\epsilon_{ab}G^{ab},\,\,\,\,L^{-1}=1-\frac{1}{2}\epsilon_{ab}G^{ab},
\end{equation}
where $G^{ab}$ are the {\em generators} of the spinor representation of the Lorentz group:
\begin{equation}
G^{ab}=\frac{1}{4}(\gamma^a\gamma^b-\gamma^b\gamma^a).
\label{gen}
\end{equation}

A {\em spinor} $\psi$ is defined to be a quantity that, under tetrad rotations, transforms according to~\cite{BJ}
\begin{equation}
\tilde{\psi}=L\psi.
\end{equation}
An {\em adjoint spinor} $\bar{\psi}$ is defined to be a quantity that transforms according to
\begin{equation}
\tilde{\bar{\psi}}=\bar{\psi}L^{-1}.
\end{equation}
Consequently, the Dirac matrices $\gamma^a$ can be regarded as quantities that have, in addition to the invariant index $a$, one spinor index and one adjoint-spinor index.
The derivative of a spinor does not transform like a spinor since
\begin{equation}
\tilde{\psi}_{,\mu}=L\psi_{,\mu}+L_{,\mu}\psi.
\end{equation}
If we introduce the {\em spinor connection} $\Gamma_\mu$ that transforms according to
\begin{equation}
\tilde{\Gamma}_\mu=L\Gamma_\mu L^{-1}+L_{,\mu}L^{-1},
\label{trans}
\end{equation}
then the {\em covariant derivative} of a spinor~\cite{Lord}:
\begin{equation}
\psi_{:\mu}=\psi_{,\mu}-\Gamma_\mu \psi,
\label{covsp}
\end{equation}
is a spinor:
\begin{equation}
\tilde{\psi}_{:\mu}=L\psi_{:\mu}.
\end{equation}
Similarly, one can show that the spinor-covariant derivative of the Dirac matrices $\gamma^a$ is
\begin{equation}
\gamma^a_{\phantom{a}:\mu}=-[\Gamma_\mu,\gamma^a]
\label{covar}
\end{equation}
since $\tilde{\gamma}^\mu=L\gamma^\mu L^{-1}$ due to Eq.~(\ref{spinor}) and $\gamma^a_{\phantom{a},\mu}=0$.

\section{Lorentz connection}

Covariant differentiation of a contravariant vector $V^\mu$ and a covariant vector $W_\mu$ in a relativistic theory of gravitation introduces the {\em affine connection} $\Gamma^{\,\,\rho}_{\mu\,\nu}$:
\begin{equation}
V^\mu_{\phantom{\mu};\nu}=V^\mu_{\phantom{\mu},\nu}+\Gamma^{\,\,\mu}_{\rho\,\nu}V^\rho,\,\,\,\,W_{\mu;\nu}=W_{\mu,\nu}-\Gamma^{\,\,\rho}_{\mu\,\nu}W_\rho,
\label{covder}
\end{equation}
where the semicolon denotes the {\em covariant derivative} with respect to coordinate indices.\footnote{
The definitions~(\ref{covder}) are related to one another via the condition $(V^\mu W_\mu)_{;\nu}=(V^\mu W_\mu)_{,\nu}$.
}
The affine connection in general relativity is constrained to be symmetric, $\Gamma^{\,\,\rho}_{\mu\,\nu}=\Gamma^{\,\,\rho}_{\nu\,\mu}$, and metric compatible, $g_{\mu\nu;\rho}=0$.
For a general spacetime we do not impose these constraints.
As a result, raising and lowering of coordinate indices does not commute with covariant differentiation with respect to $\Gamma^{\,\,\rho}_{\mu\,\nu}$.

Let us define:
\begin{equation}
\omega^\mu_{\phantom{\mu}a\nu}=e^\mu_{a;\nu}=e^\mu_{a,\nu}+\Gamma^{\,\,\mu}_{\rho\,\nu}e^\rho_a.
\label{omega}
\end{equation}
The quantities $\omega^{ab}_{\phantom{ab}\mu}=e^a_\rho \eta^{bc} \omega^\rho_{\phantom{\rho}c\mu}$ transform like tensors under coordinate transformations.
We can extend the notion of covariant differentiation to quantities with Lorentz coordinate-invariant indices by regarding $\omega^{ab}_{\phantom{ab}\mu}$ as a connection~\cite{Hehl,Lord}:
\begin{equation}
V^a_{\phantom{a}|\mu}=V^a_{\phantom{a},\mu}+\omega^{a}_{\phantom{a}b\mu}V^b.
\end{equation}
The covariant derivative of a scalar $V^a W_a$ coincides with its ordinary derivative:
\begin{equation}
(V^a W_a)_{|\mu}=(V^a W_a)_{,\mu},
\end{equation}
which gives
\begin{equation}
W_{a|\mu}=W_{a,\mu}-\omega^{b}_{\phantom{b}a\mu}W_b.
\end{equation}
We can also assume that the {\em covariant derivative} $|$ recognizes coordinate and spinor indices, acting on them like $;$ and $:$, respectively.
Accordingly, the covariant derivative of the Dirac matrices $\gamma^a$ is~\cite{LR}
\begin{equation}
\gamma^a_{\phantom{a}|\mu}=\omega^{a}_{\phantom{a}b\mu}\gamma^b-[\Gamma_\mu,\gamma^a].
\end{equation}
The definition~(\ref{omega}) can be written as~\cite{Lord}
\begin{equation}
e^\mu_{a|\nu}=e^\mu_{a,\nu}+\Gamma^{\,\,\mu}_{\rho\,\nu}e^\rho_a-\omega^b_{\phantom{b}a\nu}e^\mu_b=0.
\label{zero}
\end{equation}

Eq.~(\ref{zero}) implies that the total covariant differentiation commutes with converting between coordinate and Lorentz indices.
This equation also determines the {\em Lorentz connection} $\omega^{ab}_{\phantom{ab}\mu}$, also called the {\em spin connection}, in terms of the affine connection, tetrad and its derivatives.
Conversely, the affine connection is determined by the Lorentz connection, tetrad and its derivatives~\cite{spinor5}:
\begin{equation}
\Gamma^{\,\,\rho}_{\mu\,\nu}=\omega^\rho_{\phantom{\rho}\mu\nu}+e^a_{\mu,\nu}e^\rho_a.
\label{affine}
\end{equation}
The Cartan {\em torsion tensor} $S^\rho_{\phantom{\rho}\mu\nu}=\Gamma^{\,\,\,\rho}_{[\mu\,\nu]}$ is then
\begin{equation}
S^\rho_{\phantom{\rho}\mu\nu}=\omega^\rho_{\phantom{\rho}[\mu\nu]}+e^a_{[\mu,\nu]}e^\rho_a,
\end{equation}
from which we obtain the {\em torsion vector} $S_\mu=S^\nu_{\phantom{\nu}\mu\nu}$:
\begin{equation}
S_\mu=\omega^\nu_{\phantom{\nu}[\mu\nu]}+e^a_{[\mu,\nu]}e^\nu_a.
\end{equation}

\section{Spinor connection}

We now derive the spinor connection $\Gamma_\mu$ for a general affine connection.
Deviations of such a connection from the Levi-Civita metric-compatible connection are characterized by the {\em nonmetricity tensor}: 
\begin{equation}
N_{\mu\nu\rho}=g_{\mu\nu;\rho}.
\label{nonm0}
\end{equation}
Eq.~(\ref{nonm0}) yields:
\begin{equation}
\eta_{ab|\rho}=N_{ab\rho},\,\,\,\,\eta^{ab}_{\phantom{ab}|\rho}=-N^{ab}_{\phantom{ab}\rho},
\label{nonm1}
\end{equation}
from which it follows that
\begin{equation}
\omega_{(ab)\mu}=-\frac{1}{2}N_{ab\mu},
\label{nonm2}
\end{equation}
i.e. the Lorentz connection is antisymmetric in first two indices {\em only} for a metric-compatible affine connection~\cite{Hehl}.\footnote{
As a result, raising and lowering of Lorentz indices does not commute with covariant differentiation; it commutes only with ordinary differentiation.
}
In the presence of nonmetricity (lack of metric compatibility) the covariant derivative of the Dirac matrices deviates from zero.
From Eqs.~(\ref{anticom1}) and~(\ref{nonm1}) it follows that~\cite{spinor2}
\begin{equation}
\gamma^a_{\phantom{a}|\mu}=-\frac{1}{2}N^a_{\phantom{a}b\mu}\gamma^b.
\label{linear}
\end{equation}
Multiplying both sides of this equation by $\gamma_a$ (from the left) yields
\begin{equation}
\omega_{ab\mu}\gamma^a \gamma^b-\gamma_a \Gamma_\mu \gamma^a+4\Gamma_\mu=-\frac{1}{2}N^c_{\phantom{c}c\mu}.
\label{FI1}
\end{equation}

We seek the solution of Eq.~(\ref{FI1}) in the form:
\begin{equation}
\Gamma_\mu=-\frac{1}{4}\omega_{[ab]\mu}\gamma^a \gamma^b-A_\mu,
\label{FI2}
\end{equation}
where $A_\mu$ is a spinor quantity with one vector index.
Substituting Eq.~(\ref{FI2}) to~(\ref{FI1}) and using the identity $\gamma_c \gamma^a \gamma^b \gamma^c=4\eta^{ab}$ give
\begin{equation}
-\gamma_a A_\mu \gamma^a+4A_\mu=\frac{1}{2}N^c_{\phantom{c}c\mu}+\omega^c_{\phantom{c}c\mu}.
\label{FI3}
\end{equation}
The right-hand side of Eq.~(\ref{FI3}) vanishes because of Eq.~(\ref{nonm2}) so $A_\mu$ is simply an arbitrary vector multiple of the unit matrix~\cite{Lord,BC}.
We can write Eq.~(\ref{FI2}) as
\begin{equation}
\Gamma_\mu=-\frac{1}{4}\omega^{(A)}_{ab\mu}\gamma^a \gamma^b,
\end{equation}
where $\omega^{(A)}_{ab\mu}$ is related to $\omega_{[ab]\mu}$ by a {\em projective transformation}:\footnote{
The transformation~(\ref{project}) is related, because of Eq.~(\ref{affine}), to a projective transformation of the affine connection: $\Gamma^{\,\,\rho}_{\mu\,\nu}\rightarrow\Gamma^{\,\,\rho}_{\mu\,\nu}+\delta^\rho_\mu A_\nu$~\cite{Ein}.
}
\begin{equation}
\omega^{(A)}_{ab\mu}=\omega_{[ab]\mu}+\eta_{ab}A_\mu.
\label{project}
\end{equation}
If we assume that $A_\mu$ represents some non-gravitational field then we can, for spinors in a purely gravitational field, set $A_\mu=0$.\footnote{
The invariance of Eq.~(\ref{FI1}) under the addition of a vector multiple $A_\mu$ of the unit matrix to the spinor connection allows to introduce gauge fields interacting with spinors~\cite{BC}.
Eq.~(\ref{covsp}) can be written as $\psi_{:\mu}=(\partial_\mu+A_\mu)\psi+\frac{1}{4}\omega_{[ab]\mu}\gamma^a \gamma^b \psi$, from which it follows that if the vector $A_\mu$ is imaginary then we can treat it as a gauge field.
}
Therefore the spinor connection $\Gamma_\mu$ for a general affine connection has the form of the {\em Fock--Ivanenko coefficients} of general relativity~\cite{HD,FoIv,Lord,spinor2,Uti,FI}:
\begin{equation}
\Gamma_\mu=-\frac{1}{4}\omega_{[ab]\mu}\gamma^a \gamma^b,
\label{FI4}
\end{equation}
with the {\em antisymmetric} part of the Lorentz connection.\footnote{
The Fock--Ivanenko coefficients~(\ref{FI4}) can also be written in terms of the generators of the spinor representation of the Lorentz group~(\ref{gen}): $\Gamma_\mu=-\frac{1}{2}\omega_{ab\mu}G^{ab}$.
}
Using the definition~(\ref{omega}), we can also write Eq.~(\ref{FI4}) as
\begin{equation}
\Gamma_\mu=-\frac{1}{8}e^\nu_{c;\mu}[\gamma_\nu,\gamma^c]=\frac{1}{8}[\gamma^\nu_{\phantom{\nu};\mu},\gamma_\nu].
\end{equation}

\section{Curvature spinor}

The commutator of the covariant derivatives of a vector with respect to the affine connection defines the {\em curvature tensor} $R^\rho_{\phantom{\rho}\sigma\mu\nu}=\Gamma^{\,\,\rho}_{\sigma\,\nu,\mu}-\Gamma^{\,\,\rho}_{\sigma\,\mu,\nu}+\Gamma^{\,\,\kappa}_{\sigma\,\nu}\Gamma^{\,\,\rho}_{\kappa\,\mu}-\Gamma^{\,\,\kappa}_{\sigma\,\mu}\Gamma^{\,\,\rho}_{\kappa\,\nu}$:\footnote{
The curvature tensor can also be defined through a parallel displacement $\delta V=(V_{,\mu}-V_{;\mu})dx^\mu$ of a vector $V$ along the boundary of an infinitesimal surface element $\Delta f^{\mu\nu}$~\cite{Scho}: $\oint\delta V^\rho=-\frac{1}{2}R^\rho_{\phantom{\rho}\sigma\mu\nu}V^\sigma\Delta f^{\mu\nu}$ and $\oint\delta V_\rho=\frac{1}{2}R^\sigma_{\phantom{\sigma}\rho\mu\nu}V_\sigma\Delta f^{\mu\nu}$.
These equations are related to one another since the parallel displacement of a scalar along a closed curve vanishes.
}
\begin{equation}
V^\rho_{\phantom{\rho};\nu\mu}-V^\rho_{\phantom{\rho};\mu\nu}=R^\rho_{\phantom{\rho}\sigma\mu\nu}V^\sigma+2S^\sigma_{\phantom{\sigma}\mu\nu}V^\rho_{\phantom{\rho};\sigma},\,\,\,\,V_{\rho;\nu\mu}-V_{\rho;\mu\nu}=-R^\sigma_{\phantom{\sigma}\rho\mu\nu}V_\sigma+2S^\sigma_{\phantom{\sigma}\mu\nu}V_{\rho;\sigma}.
\end{equation}
In analogous fashion, the commutator of the total covariant derivatives of a spinor:
\begin{equation}
\psi_{|\nu\mu}-\psi_{|\mu\nu}=K_{\mu\nu}\psi+2S^\rho_{\phantom{\rho}\mu\nu}\psi_{|\rho},
\end{equation}
defines the antisymmetric {\em curvature spinor} $K_{\mu\nu}$~\cite{Lord}:
\begin{equation}
K_{\mu\nu}=\Gamma_{\mu,\nu}-\Gamma_{\nu,\mu}+[\Gamma_\mu,\Gamma_\nu].
\end{equation}
From Eq.~(\ref{trans}) it follows that $K_{\mu\nu}$ transforms under tetrad rotations like the Dirac matrices $\gamma^a$ (with one spinor index and one adjoint-spinor index):
\begin{equation}
\tilde{K}_{\mu\nu}=LK_{\mu\nu}L^{-1}.
\end{equation}

Eq.~(\ref{linear}) is equivalent to
\begin{equation}
\gamma^\rho_{\phantom{\rho}|\mu}=-\frac{1}{2}N^\rho_{\phantom{\rho}\sigma\mu}\gamma^\sigma.
\end{equation}
The commutator of the covariant derivatives of the spacetime-dependent Dirac matrices with respect to the affine connection is then:
\begin{equation}
2\gamma^\rho_{\phantom{\rho}|[\nu\mu]}=-(N^\rho_{\phantom{\rho}\sigma[\nu}\gamma^\sigma)_{|\mu]}.
\end{equation}
Multiplying both sides of this equation by $\gamma_\rho$ (from the left) and using
\begin{equation}
2\gamma^\rho_{\phantom{\rho}|[\nu\mu]}=R^\rho_{\phantom{\rho}\sigma\mu\nu}\gamma^\sigma+2S^\sigma_{\phantom{\sigma}\mu\nu}\gamma^\rho_{\phantom{\rho}|\sigma}+[K_{\mu\nu},\gamma^\rho]
\end{equation}
and Eq.~(\ref{anticom2}) yield
\begin{eqnarray}
& & R_{\rho\sigma\mu\nu}\gamma^\rho \gamma^\sigma-S^\sigma_{\phantom{\sigma}\mu\nu}N^\rho_{\phantom{\rho}\rho\sigma}+\gamma_\rho K_{\mu\nu}\gamma^\rho-4K_{\mu\nu}=-\gamma_\rho(N^\rho_{\phantom{\rho}\sigma[\nu}\gamma^\sigma)_{|\mu]} \nonumber \\
& & =-N^\rho_{\phantom{\rho}\rho[\nu;\mu]}+\frac{1}{2}N_{\rho\sigma[\nu}N^{\lambda\rho}_{\phantom{\sigma\rho}\mu]}\gamma_\lambda \gamma^\sigma=-N^\rho_{\phantom{\rho}\rho[\nu,\mu]}-S^\sigma_{\phantom{\sigma}\mu\nu}N^\rho_{\phantom{\rho}\rho\sigma}+\frac{1}{2}N_{\rho\sigma[\nu}N^{\rho\lambda}_{\phantom{\rho\lambda}\mu]}\gamma_\lambda \gamma^\sigma.
\label{cs1}
\end{eqnarray}

We seek the solution of Eq.~(\ref{cs1}) in the form:
\begin{equation}
K_{\mu\nu}=\frac{1}{4}R_{[\rho\sigma]\mu\nu}\gamma^\rho \gamma^\sigma-\frac{1}{8}N_{\rho\sigma[\nu}N^{\rho\lambda}_{\phantom{\rho\lambda}\mu]}\gamma_\lambda \gamma^\sigma+B_{\mu\nu},
\label{cs2}
\end{equation}
where $B_{\mu\nu}$ is a spinor quantity with two vector indices.
Substituting Eq.~(\ref{cs2}) to~(\ref{cs1}) gives
\begin{equation}
\gamma_\rho B_{\mu\nu}\gamma^\rho-4B_{\mu\nu}=-Q_{\mu\nu}-N^\rho_{\phantom{\rho}\rho[\nu,\mu]},
\label{cs3}
\end{equation}
where
\begin{equation}
Q_{\mu\nu}=R_{\rho\sigma\mu\nu}g^{\rho\sigma}=\Gamma^{\,\,\rho}_{\rho\,\nu,\mu}-\Gamma^{\,\,\rho}_{\rho\,\mu,\nu}
\label{second}
\end{equation}
is the {\em second Ricci tensor}, also called the tensor of {\em homothetic curvature}~\cite{spinor1} or the {\em segmental curvature tensor}~\cite{spinor2}.
The right-hand side of Eq.~(\ref{cs3}) vanishes because of Eq.~(\ref{nonm0}) so $B_{\mu\nu}$ is simply an antisymmetric-tensor multiple of the unit matrix.
The tensor $B_{\mu\nu}$ is related to the vector $A_\mu$ in Eq.~(\ref{FI2}) by\footnote{
If $\psi$ is a pure spinor, i.e. has no non-spinor indices, then $A_\mu$ is a pure vector, like the electromagnetic potential~\cite{BC}, and $[A_\mu,A_\nu]=0$.
If $\psi$ has non-spinor indices corresponding to some symmetries, e.g., the electron--neutrino symmetry, then $A_\mu$ is also assigned these indices and $[A_\mu,A_\nu]$ can be different from zero.
}
\begin{equation}
B_{\mu\nu}=A_{\nu,\mu}-A_{\mu,\nu}+[A_\mu,A_\nu].
\label{gauge}
\end{equation}
Setting $A_\mu=0$, which corresponds to the absence of non-gravitational fields, yields $B_{\mu\nu}=0$.\footnote{
Eq.~(\ref{gauge}) resembles the definition of a field strength in terms of a field potential for a non-commutative gauge field.
The invariance of Eq.~(\ref{cs1}) under the addition of an antisymmetric-tensor multiple $B_{\mu\nu}$ of the unit matrix to the curvature spinor is related to the invariance of Eq.~(\ref{FI1}) under the addition of a vector multiple $A_\mu$ of the unit matrix to the spinor connection.
}
Therefore the curvature spinor for a general affine connection is:\footnote{
The curvature spinor~(\ref{cs4}) can also be written in terms of the generators of the spinor representation of the Lorentz group~(\ref{gen}): $K_{\mu\nu}=\frac{1}{2}R_{\rho\sigma\mu\nu}G^{\rho\sigma}-\frac{1}{4}N_{\rho\lambda\mu}N^{\rho\sigma}_{\phantom{\rho\sigma}\nu}G^\lambda_{\phantom{\lambda}\sigma}$.
}
\begin{equation}
K_{\mu\nu}=\frac{1}{4}R_{[\rho\sigma]\mu\nu}\gamma^\rho \gamma^\sigma-\frac{1}{8}N_{\rho\lambda[\mu}N^{\rho\sigma}_{\phantom{\rho\sigma}\nu]}\gamma^\lambda \gamma_\sigma.
\label{cs4}
\end{equation}

\section{Curvature and Ricci tensors}

Finally, we examine the curvature tensor for a general affine connection.
The commutator of the covariant derivatives of a tetrad with respect to the affine connection is
\begin{equation}
2e^\rho_{a;[\nu\mu]}=R^\rho_{\phantom{\rho}\sigma\mu\nu}e^\sigma_a+2S^\sigma_{\phantom{\sigma}\mu\nu}e^\rho_{a;\sigma}.
\end{equation}
This commutator can also be expressed in terms of the Lorentz connection:
\begin{equation}
e^\rho_{a;[\nu\mu]}=\omega^\rho_{\phantom{\rho}a[\nu;\mu]}=(e^\rho_b \omega^b_{\phantom{b}a[\nu})_{;\mu]}=\omega_{ba[\nu}\omega^{\rho b}_{\phantom{\rho b}\mu]}+\omega^b_{\phantom{b}a[\nu;\mu]}e^\rho_b=\omega_{ba[\nu}\omega^{\rho b}_{\phantom{\rho b}\mu]}+\omega^b_{\phantom{b}a[\nu,\mu]}e^\rho_b+S^\sigma_{\phantom{\sigma}\mu\nu}\omega^\rho_{\phantom{\rho}a\sigma}.
\end{equation}
Consequently, the curvature tensor with two Lorentz and two coordinate indices depends only on the Lorentz connection and its first derivatives~\cite{KS,Lord,Uti}:
\begin{equation}
R^{ab}_{\phantom{ab}\mu\nu}=\omega^{ab}_{\phantom{ab}\nu,\mu}-\omega^{ab}_{\phantom{ab}\mu,\nu}+\omega^a_{\phantom{a}c\mu}\omega^{cb}_{\phantom{cb}\nu}-\omega^a_{\phantom{a}c\nu}\omega^{cb}_{\phantom{cb}\mu}.
\label{curva}
\end{equation}

The contraction of the tensor~(\ref{curva}) with a tetrad leads to two {\em Lorentz-connection Ricci tensors}:
\begin{eqnarray}
& & R^a_\mu=R^{[ab]}_{\phantom{[ab]}\mu\nu}e^\nu_b, \\
& & P^a_\mu=R^{(ab)}_{\phantom{(ab)}\mu\nu}e^\nu_b,
\end{eqnarray}
that are related to the sum and difference, respectively, of the {\em Ricci tensor}, $R_{\mu\nu}=R^\rho_{\phantom{\rho}\mu\rho\nu}$, and the tensor $C_{\mu\nu}=R_{\mu\rho\nu\sigma}g^{\rho\sigma}$.
The contraction of the tensor $R^a_\mu$ with a tetrad gives the {\em Ricci scalar},
\begin{equation}
R=R^a_\mu e^\mu_a=R^{ab}_{\phantom{ab}\mu\nu}e^\mu_a e^\nu_b,
\end{equation}
while the contraction of $P^a_\mu$ gives zero.
The contraction of the curvature tensor~(\ref{curva}) with the Lorentz metric tensor $\eta_{ab}$ gives the second Ricci tensor~(\ref{second}),\footnote{
If the Lorentz connection is antisymmetric, the curvature tensor~(\ref{curva}) is antisymmetric in the Lorentz indices and the tensors $P^a_\mu$ and $Q_{\mu\nu}$ vanish.
}
\begin{equation}
Q_{\mu\nu}=\omega^c_{\phantom{c}c\nu,\mu}-\omega^c_{\phantom{c}c\mu,\nu}.
\end{equation}

Under the projective transformation~(\ref{project}) the tensor~(\ref{curva}) changes according to
\begin{equation}
R^{ab}_{\phantom{ab}\mu\nu}\rightarrow R^{ab}_{\phantom{ab}\mu\nu}+\eta^{ab}B_{\mu\nu},
\end{equation}
where $B_{\mu\nu}$ is given by Eq.~(\ref{gauge}), and so does $R^{(ab)}_{\phantom{(ab)}\mu\nu}$.
Consequently, the tensor $P^a_\mu$ changes according to
\begin{equation}
P^a_\mu\rightarrow P^a_\mu+e^{a\nu}B_{\mu\nu}.
\label{proj1}
\end{equation}
The tensor $R^{[ab]}_{\phantom{[ab]}\mu\nu}$ is projectively invariant and so are $R^a_\mu$ and $R$.\footnote{
The Einstein--Cartan--Kibble--Sciama formulation of gravitation~\cite{Hehl,KS}, where the tetrad and Lorentz connection are dynamical variables, is based on the Lagrangian density $\textgoth{L}=\textgoth{e}R$.
}
The second Ricci tensor changes according to
\begin{equation}
Q_{\mu\nu}\rightarrow Q_{\mu\nu}+8A_{[\nu,\mu]},
\label{proj2}
\end{equation}
so it is invariant only under {\em special projective transformations}~\cite{Ein} with $A_\mu=\lambda_{,\mu}$, where $\lambda$ is a scalar.
If $[A_\mu,A_\nu]=0$ then Eqs.~(\ref{proj1}) and~(\ref{proj2}) yield the projective invariance of the tensor:
\begin{equation}
I^a_\mu=4P^a_\mu-Q_{\mu\nu}e^{a\nu}.
\end{equation}



\end{document}